# Laguerre asymptotic formula and stability of Landau levels influenced by an electric field

J. Chee  and  Y.-H. Zhang

Department of Physics

Tianjin Polytechnic University

Tianjin, China

**Abstract:** Consider the quantum evolution of a charged particle subjected to a uniform magnetic field and an electric field $\vec{E}(t)$ that exists for a finite period of time. The electric field can induce intra-Landau level transitions (magnetic translations) that do not change the energy of the particle. It may also induce energy changing inter-Landau level transitions. Our purpose in this paper is two-fold: We first demonstrate that the inter-Landau level transition probability is completely determined by the Fourier component of the electric field corresponding to the cyclotron frequency. Then we point out that the Fejer asymptotic form of Laguerre polynomials implies that no matter how small the Fourier component is, inter-Landau level transition probability from a fixed Landau level to other energy levels can be arbitrarily close to 1 if the original Landau energy level is high enough, i.e., influenced by a given electric field, Landau levels of higher energy are less stable asymptotically and their transition probabilities are explicitly predicted by the Fejer formula.

## 1. Introduction

The introduction of the Berry phase concept [1] in quantum mechanics has shed new light on the time evolution of driven systems. For instance, the factorization of the time evolution operator in the spirit of the geometric phase can be a useful tool in obtaining information on nonadiabatic transitions for time-dependent systems [2].

In this paper our purpose is to study in detail the nonadiabatic transition probabilities for the Landau level problem with a general time dependent electric field $\vec{E}(t)$ in a plane perpendicular to the magnetic field. The calculation of nonadiabatic transitions is an important aspect of quantum physics which has many consequences [3].

The Hamiltonian for such a system is the sum of the usual Landau Hamiltonian $H_0$ and the potential energy of the charge in the electric field:

$$H = H_0 + V(x,t),$$
$$= \frac{1}{2m}[\vec{p} - \frac{q}{c}\vec{A}(x)]^2 - qE_1(t)x_1 - qE_2(t)x_2, \quad (1)$$

where the magnetic field $\vec{B} = B\vec{e}_z$. The vector potential $\vec{A}(x)$ satisfies

$$A_x(\vec{x}) = -\frac{B}{2}y, \quad A_y(\vec{x}) = \frac{B}{2}x, \quad A_z(\vec{x}) = 0.$$

The electric field $\vec{E}(t)$ is a general time-dependent electric field

$$\vec{E}(t) = E_1(t)\vec{e}_1 + E_2(t)\vec{e}_2.$$

We observe that the problem has been studied before from the perspective of constructing the quantum propagator [4, 5]. The propagator, however, does not separate the intra landau level transitions, which do not change the energy level and the inter-Landau level transitions, which is the focus of this paper. The crucial starting point of this paper is that the Berry phase concept lends itself in constructing a unique factorization of the time evolution operator that can be used to give a simple expression for the transition probability among different energy levels. The explicit construction of time evolution operator has many consequences for driven systems and for instance, in studying the experimental signature of the Berry phase, Tycko [6] pointed out that the

Abelian part of the geometric phase has manifestations in nuclear quadrupole resonance by studying how the existence of the Berry phase phenomenon affects the time evolution operator.

In a previous work [7], we showed that the time evolution of the Landau charged particle in a general time dependent electric field can be naturally factorized into three factors: a geometric operator $M(C_R)$, a dynamical operator $D(t)$ and a nonadiabatic operator $J(C_u)$:

$$U(t,0) = M(C_R)D(t)J(C_u), \quad (2)$$

where $D(t) = \exp(-iH_0 t/\hbar)$ is the evolution operator generated by the usual Landau Hamiltonian;

$$M(C_R) = e^{i\beta(C_R)}\exp(-i\omega_\mu R_\mu/\hbar)$$ describes a path-ordered magnetic translation [8, 9] corresponding to the path $C_R$ in parameter space traversed by

$$\vec{R}(t) = \frac{c}{B}\int_0^t (E_2(s)\vec{e}_1 - E_1(s)\vec{e}_2)\,ds.$$

$\vec{R}(t)$ represents the global shift in position of the wave packet due to the electric field, and where

$$\beta(C_{\vec{R}}) = -\frac{qB}{\hbar c}\vec{e}_3 \cdot \frac{1}{2}\int_{C_R}\vec{R}\times d\vec{R} = -\frac{qB}{\hbar c}S$$

is determined by the path $C_R$ traversed by $\vec{R}(t)$.

$J(C_u)$ is an operator which determines the transition probabilities among the Landau levels:

$$J(C_u) = e^{i\gamma(C_u)}\exp(\pi u/\hbar - \pi^+ u^*/\hbar), \quad (3)$$

where $\gamma(C_u) = -\frac{qB}{\hbar c}4S(C_u)$, $S(C_u)$ is the area enclosed by the path traversed by $u(t)$ in a complex $u$-plane and the straight line connecting the end and initial points of the path. The path $C_u$ is traversed by the complex parameter $u$. And the parameter $u$ that determines the operator $J(C_u)$ is given by

$$u = \frac{i}{2}\int_0^t e^{-i\omega s}\frac{d}{ds}R^*(s)\,ds = \frac{-c}{2B}\int_0^t e^{-i\omega s}E^*(s)\,ds, (4)$$ So the parameter $u$ is determined by the function of the electric field $\vec{E}(t)$. $\pi^+$ and $\pi$ are proportional to the raising and lowering operators. Note that if the electric field exists for only a finite period of time between 0 and t, then Eq. (4) is the Fourier transform of $E^*(s)$, therefore we conclude that for an electric field that exists for a finite period of time inter-Landau level transitions is completely determined by the time domain Fourier transform of the electric field.

We note in passing that the above factorization has implications for the quantum adiabatic theorem [10, 11] involving infinitely degenerate energy levels because we can recast the Hamiltonian (1) into gauge equivalent Hamiltonian through gauge transformation:

$$H_L = \frac{1}{2m}\left[p - \frac{q}{c}A_L(x,R)\right]^2, \quad (5)$$

$$\psi_L(x,t) = \exp\left[-i\frac{q}{\hbar c}\chi(x,R)\right]\psi(x,t), \quad (6)$$

$$\vec{A}_L(\vec{x},\vec{R}) = \vec{A}(\vec{x}) - \nabla\chi(\vec{x},\vec{R}) = \vec{A}(\vec{x}) + B\vec{e}_3\times\vec{R}(t), (7)$$

where $\chi(x,t) = -BR_2(t)x_1 + BR_1(t)x_2$, and the time-evolution operators $U(t,0)$ are related by

$$U_L(t,0) = \exp[-i\frac{q}{\hbar c}\chi(\vec{x},\vec{R})]\,U(t,0) \quad (8)$$

Note that the Hamiltonian $H_L(t)$, unlike the gauge equivalent $H(t)$, has energy eigenvalues $E_n = \hbar\omega\left(n+\frac{1}{2}\right)$, where $\omega = qB/mc$ is the cyclotron frequency. Therefore, it is in the gauge of $H_L(t)$ that $D(t)$ carries the dynamical phase factor of the eigenstates of the Hamiltonian. This should relate to the quantum adiabatic theorem, where $G(C_{\vec{R}}) = \exp[-i\frac{q}{\hbar c}\chi(\vec{x},\vec{R})]M(C_{\vec{R}})$ is a geometrical operator completely determined by the path of $\vec{R}(t)$ which brings an initial eigenstate of $H_L(0)$ to an instantaneous eigenstate of $H_L(t)$. $J(C_u)$ is the exponential of explicit functions of the canonical variables, it thus provides an explicit example of the quantum adiabatic theorem involving infinitely degenerate energy levels. The factorization separates the effect caused by the electric field into a geometric operator and a nonadiabatic operator, therefore it makes possible to calculate nonadiabatic transition probabilities among different energy levels.

## 2. Laguerre polynomial and nonadiabatic transitions

In this paper, our focus of attention is on applying the above factorization to study in detail nonadiabatic transitions induced by the electric field. Since $M(C_R)D(t)$ in the time evolution operator does not cause transitions among Landau levels, the transitions are caused by the action of $J(C_u)$ on $|\Phi(n,0)\rangle$ (which is the initial state that belongs to the n-th Landau level) only. In the most general case, the expression

$$|\langle \Phi(m,0)| J(C_u(t)) |\Phi(n,0)\rangle|^2$$

needs to be evaluated in order to determine the transition probability from an initial state (at t = 0) that is at the n-th Landau energy level to an m-th energy level at time t. Using the formula $e^{A+B} = e^A e^B e^{-\frac{1}{2}[A,B]}$ and the commutation relation $[a,a^+]=1$, $(\pi/\hbar=ka)$ we have another expression of (3):

$$J(C_u) = e^{i\gamma(C_u)} e^{-\frac{1}{2}|uk|^2} e^{-u^*ka^+} e^{uka}, \quad (9)$$

Therefore, we have the following general expression for the matrix elements:

$$\langle m | J(C_u) | n \rangle = e^{i\gamma(C_u)} e^{-\frac{1}{2}|uk|^2} (e^{-uka} | m \rangle)^+ (e^{uka} | n \rangle), \quad (10)$$

which represents mixing of the Landau energy levels for the most general type of the electric field which is not necessarily small and which does not have to change slowly. In particular, this formula implies that for the ground state we have the following matrix element:

$$\langle 0 | J(C_u) | 0 \rangle = e^{i\gamma(C_u)} e^{-\frac{1}{2}|uk|^2}, \quad (11)$$

which implies that if we start with a ground state of $H_0$ at time 0, the probability that it remains to be in a ground state is $|\langle 0|J(C_u(t))|0\rangle|^2 = e^{-|uk|^2}$. In general the transition probability from an initial state (at t = 0) that is at the n-th Landau energy level to all the other energy levels at time t is given by:

$$\sum_{m \neq n} |\langle m | J(C_u) | n \rangle|^2 = 1 - |\langle n | J(C_u) | n \rangle|^2, \quad (12)$$

Using the relation $a|n\rangle = \sqrt{n}|n-1\rangle$, we observe that the expression for $\langle n| J(C_u) |n\rangle$ can be calculated in a simple way as follows:

$$\begin{aligned}
& e^{-i\gamma(C_u)} \langle n| J(C_u) |n\rangle \\
&= e^{-\frac{1}{2}|uk|^2} (e^{-uka}|n\rangle)^+ (e^{uka}|n\rangle) \\
&= e^{-\frac{1}{2}x} (e^{-uka}|n\rangle)^+ (e^{uka}|n\rangle) \\
&= e^{-\frac{1}{2}x} \left( \langle n| + u^*k\sqrt{n}\langle n-1| + \frac{(u^*k)^2}{2!}\sqrt{n(n-1)}\langle n-2| + \cdots + \frac{(u^*k)^n}{n!}\sqrt{n!}\langle 0| \right) \\
&\quad \times \left( |n\rangle + uk\sqrt{n}|n-1\rangle + \frac{(uk)^2}{2!}\sqrt{n(n-1)}|n-2\rangle + \cdots + \frac{(uk)^n}{n!}\sqrt{n!}|0\rangle \right) \\
&= e^{-\frac{1}{2}x} \left( C_n^0 + (-1) C_n^1 x + \frac{1}{2!} C_n^2 x^2 \cdots + (-1)^n \frac{1}{n!} C_n^n x^n \right) \\
&= e^{-\frac{1}{2}x} L_n(x) \quad (13)
\end{aligned}$$

where we have used one of the definitions of the Laguerre polynomial and where we have $x = |uk|^2$.

This shows that the amplitude for a certain Landau energy level to remain at that energy level under the influence of the electric field is proportional to the Laguerre polynomial in terms of the time domain Fourier transform of the electric field. Although the connection between the matrix elements of the coherent state displacement operator and the Lagurre polynomial has been known in the literature [12], as far as we know, this relation has not been linked to the transition probability of the Landau electron subject to a time dependent electric field.

Furthermore we want to emphasize here that if we make use of the Fejer asymptotic formula [13] of the Laguerre polynomial:

$$L_n(x) \to \frac{e^{\frac{x}{2}}}{\sqrt{\pi}(x)^{1/4}(n+1)^{1/4}} \times \cos\left(2\sqrt{(n+1)x} - \pi/4\right) + O\left((n+1)^{-3/4}\right) \quad (14)$$

which applies for fixed x and for large n, we would get the following

$$e^{-i\gamma(C_u)}\langle n| J(C_u) |n\rangle = \frac{1}{\sqrt{\pi}(x)^{1/4}(n+1)^{1/4}} \times \cos\left(2\sqrt{(n+1)x} - \pi/4\right) + O\left((n+1)^{-3/4}\right) \quad (15)$$

where $O((n+1)^{-3/4})$ is of the order $(n+1)^{-3/4}$ for large $n$. From this it is clear that the transition probability from energy level n to all other energy levels, which is equal to $1 - |\langle n|J(C_u)|n\rangle|^2$ can be arbitrarily close to 1 as long as n is big enough. That is to say, no matter how small the magnitude of the Fourier component $|u|$ is, there will be significant transitions as long as the energy level is high enough. This is a key observation of this paper. This conclusion applies for the most general form of $\vec{E}(t)$, it is a result that goes beyond perturbation theory. In particular, the electric field does not have to be a continuous function, which includes the case where $\vec{E}(t)$ is a noisy background.

We now study numerically how the transition probability varies as function of n and x, respectively.

In Figure 1 we have plotted how the transition probability varies with energy level n when the Fourier component, or x is fixed at x=8. We let n vary from n = 0 to n = 160. It demonstrates how the transition probability grows and eventually tends to 1 as energy level increases.

In Figure2, we have plotted the dependence of transition probability on x when the energy level $n$ is fixed at $n = 100$. We see that as x increases, transition probability oscillates and tends to 1 for sufficiently large x.

We wish to comment here that although the oscillations as seen in Fig.1 and 2 are reminiscent of similar effects in other more involved contexts such as the Shubnikov - de Haas and Weiss oscillations much studied in solid state physics [14], which may be understood as the commensurability effects between the cyclotron energy and the frequency of the electric field, and to the increase of the cyclotron radius with increasing energy, our main point in this paper is that higher Landau levels are asymptotically less stable under the perturbations of a time dependent electric field. The fact that the Landau levels are infinitely degenerate makes traditional perturbation theory not directly applicable to the calculation of nonadiabatic transitions. It is only through the Berry phase type factorization of time evolution, that such a conclusion is reached.

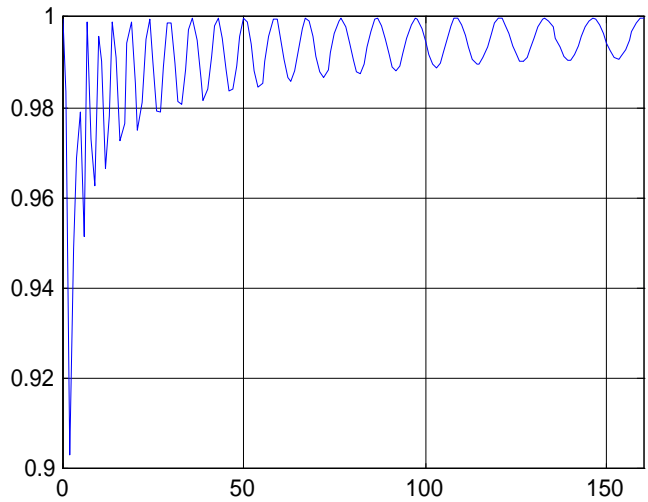

**Figure1.  Dependence of transition probability on the energy level n when x is fixed (x=10)**

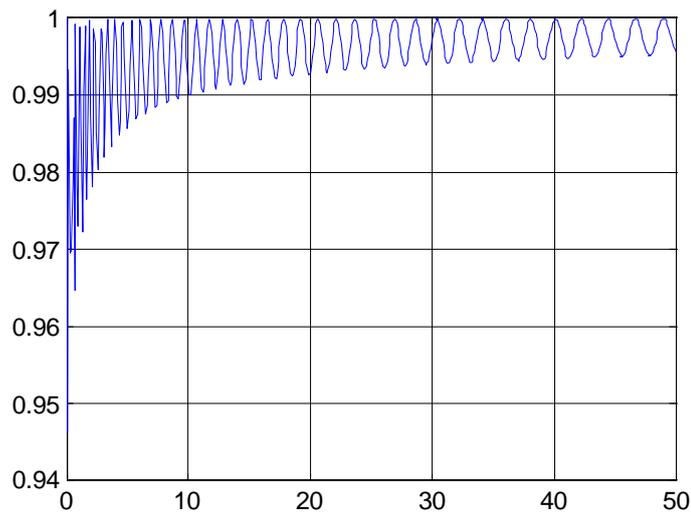

**Figure2. Dependence of transition probability on x when energy levels is field (n=100)**

### 3. Conclusions

In this paper, we studied the nonadiabatic transition probability of a Landau level due to the existence of an electric field. The electric field is quite general in its time dependence, in particular, it needs not to be a smooth function of time. Our result therefore applies, for example when the electric field is a noisy background. We point out the relevance of the Fourier transform of the electric field, the Laguerre polynomial and the Fejer assymtotic expression of Laguerre polynomial in determining the transition probability of a Landau level to all other Landau energy levels. Our main conclusion is that the nonadiabatic transition probability is energy level dependent and is described by the Laguerre polynomial. We further have made the observation, based on the Fejer asymptotic form of the Laguerre polynomial that   nonadiabatic transitions can be arbitrarily close to 1 if the Landau energy level is high enough, irrespective of how small the magnitude of the Fourier component of the electric field is, i.e., influenced by a given electric field, Landau levels of higher energy are less stable asymptotically and whose transition behavior is directly related with the Fejer asymptotic formula of the Laguerre polynomials.

An obvious question that may deserve further investigation is that for a non-zero Fourier component of the electric field, whether a high enough Landau energy level tends to go to lower energy levels or go to still higher energy levels. This involves detailed analysis the the off diagonal matrix elements of the coherent displacement operator that is directly related to the associated Laguerre polynomials.